\documentclass [preprint]{revtex4-1} 
\usepackage[utf8]{inputenc}
\usepackage{amssymb}
\usepackage{graphicx}
\usepackage{bm}
\usepackage{color}
\usepackage{textcomp}
\usepackage{amsmath}
\usepackage{epstopdf}
\usepackage{natbib}
\usepackage{units}
\usepackage{ulem}

\usepackage[colorlinks=true,citecolor=blue]{hyperref}

\newcommand{\Tr}[1]{\ensuremath{\mathrm{Tr}(#1)}}

\begin{document}

\title{ Three-Terminal Energy Harvester with Coupled Quantum Dots\\}

\author{Holger~Thierschmann$^{1,5,*}$, Rafael~S\'anchez$^2$, Bj\"orn~Sothmann$^3$, Fabian~Arnold$^1$, Christian~Heyn$^4$, Wolfgang~Hansen$^4$, Hartmut~Buhmann$^1$ and Laurens~W.~Molenkamp$^1$}
 \email{h.r.thierschmann@tudelft.nl, laurens.molenkamp@physik.uni-wuerzburg.de }

\affiliation{$^1$Physikalisches Institut (EP3), Universit\"at W\"urzburg, Am Hubland, D-97074, W\"urzburg, Germany\\
$^2$Instituto de Ciencia de Materiales de Madrid, CSIC, Cantoblanco, 28049 Madrid, Spain\\
$^3$D\'epartement de Physique Th\'eorique, Universit\'e de Gen\`eve, CH-1211 Gen\`eve 4, Switzerland\\
$^4$Institute of Applied Physics, University of Hamburg, Jungiusstrasse 11, D-20355 Hamburg, Germany\\
$^5$\text{present address:} KavliInstitute of Nanoscience, Faculty of Applied Physics, Delft University of Technology, Lorentzweg 1, 2628 CJ Delft, The Netherlands\\
\\
Nature Nanotechnology \textbf{10} 854-858 (2015) 
doi:10.1038/nnano.2015.176}

\url{http://www.nature.com/nnano/journal/v10/n10/abs/nnano.2015.176.html}\\

\begin{abstract}
	\textbf{
		Rectification of thermal fluctuations in mesoscopic conductors is the key idea of today's attempts to build nanoscale thermoelectric energy harvesters in order to convert heat into a useful electric power~\cite{White_Beyond_2008, Mahan_Best_1996, Radousky_Energy_2012}. So far, most concepts make use of the Seebeck effect in a two-terminal geometry \cite{Shakouri_Recent_2011, Humphrey_Reversible_2005, Hicks_Effect_1993, Whitney_Nonlinear_2013, Juergens_Thermoelectric_2013} where heat and charge are both carried by the same particles.
		Here, we experimentally demonstrate the working principle of a new kind of energy harvester, proposed recently \cite{Sanchez_Optimal_2011}, using two capacitively coupled quantum dots (QD). We show that due to its novel three-terminal design which spatially separates the heat reservoir from the conductor circuit, the directions of charge and heat flow become decoupled in our device. This enables us to manipulate the direction of the generated charge current by means of external gate voltages while leaving the direction of heat flow unaffected. Our results pave the way for a new generation of multi-terminal, highly efficient nanoscale heat engines.}
\end{abstract}


\maketitle

Conventional thermoelectric devices are based on the Seebeck effect in a two-terminal geometry: For two electronic reservoirs being at different temperature, the resulting heat flow $J$ is accompanied by a net charge current $I$ if transport is energy dependent, i.e. if particle-hole-symmetry is broken \cite{Cutler_Observation_1969, Sivan_Multichannel_1986, Beenakker_Theory_1992, Molenkamp_Quantum_1990}. $I$ and $J$ are carried by the same particles and thus, the direction of $I$ is directly imposed by the temperature bias.   
For applications, however, this has considerable drawbacks because the heat reservoir necessarily is part of the electrical circuit which poses the problem of good thermal insulation while maintaining good electrical conduction \cite{Shakouri_Recent_2011}.
In this context, three-terminal thermoelectrics have attracted increasing attention. These devices open up the possibility to break the intimate coupling between $I$ and $J$ by spatially separating the heat reservoir and the conductor circuit and at the same time perform at high efficiencies~\cite{Entin-Wohlman_Three_2010, Sanchez_Thermoelectric_2011, Jiang_Thermoelectric_2012, Sothmann_Rectification_2012, Jordan_Powerful_2013, Bergenfeldt_Hybrid_2014, Brandner_Strong_2013}. 

\begin{figure}[]
	\centering
	\includegraphics[width=0.7\linewidth]{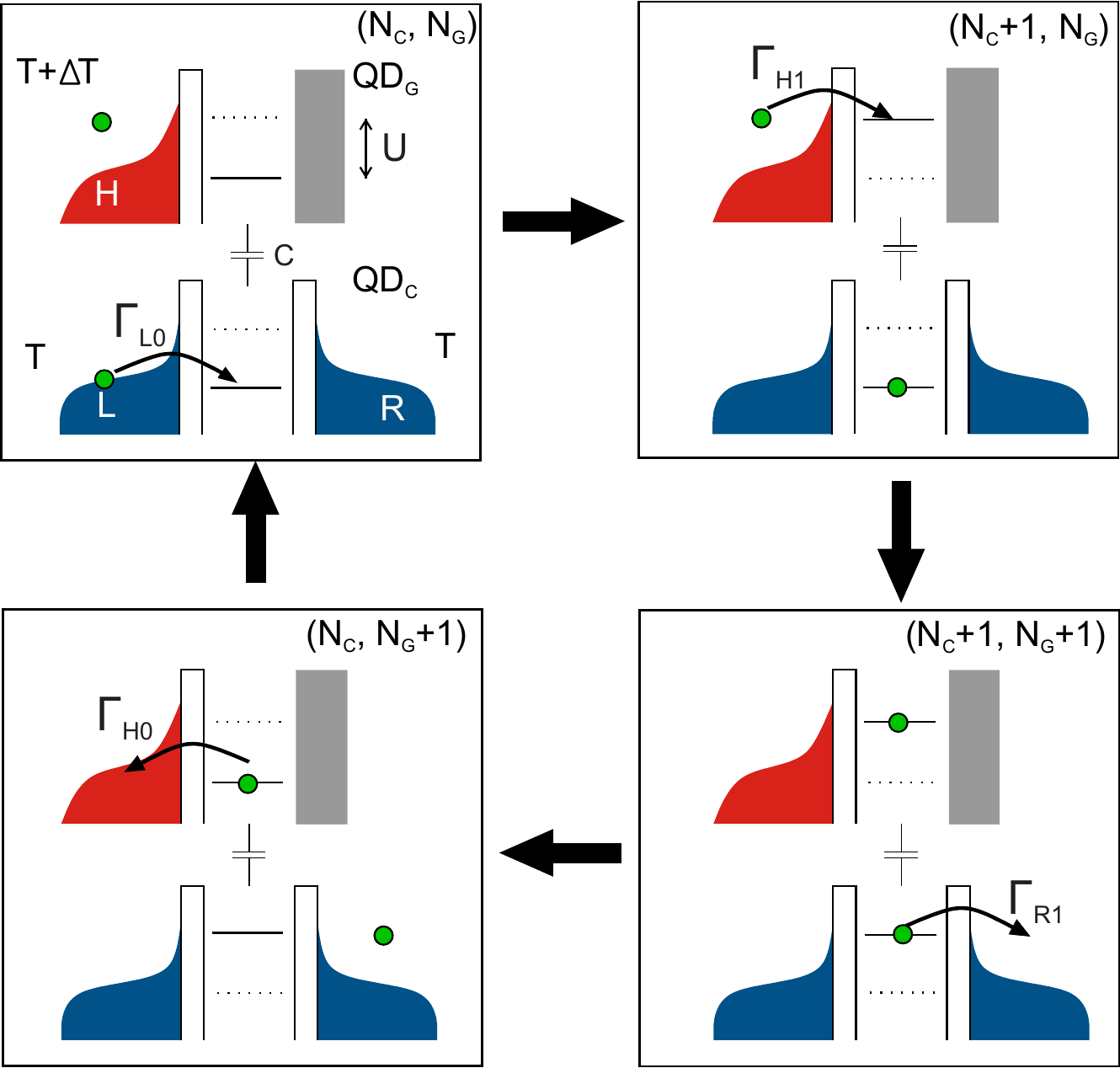}
	\caption{ Working mechanism of the energy harvester. Two quantum dots are capacitively coupled thus exchanging energy in packages of $U$, but not particles. One of them (QD$_\text{C}$) is connected to two terminals L and R. The other one (QD$_\text{G}$) is coupled to a third terminal H which is at a higher temperature. When charge fluctuations according to the depicted 4-stage sequence take place, an energy package is extracted from reservoir H and gets delivered to the cold subsystem. There, these fluctuations are rectified and converted into a charge current when the product of tunneling coefficients $\Gamma_{\rm L0}\Gamma_{\rm R1}$ differs from that of the opposite process $\Gamma_{\rm R0}\Gamma_{\rm L1}$ (\textit{not shown}), i.e. when both particle-hole-symmetry and left-right-symmetry are broken.}
	\label{fig:1}
\end{figure}

\begin{figure}[]
	\centering
	\includegraphics[width=0.6\linewidth]{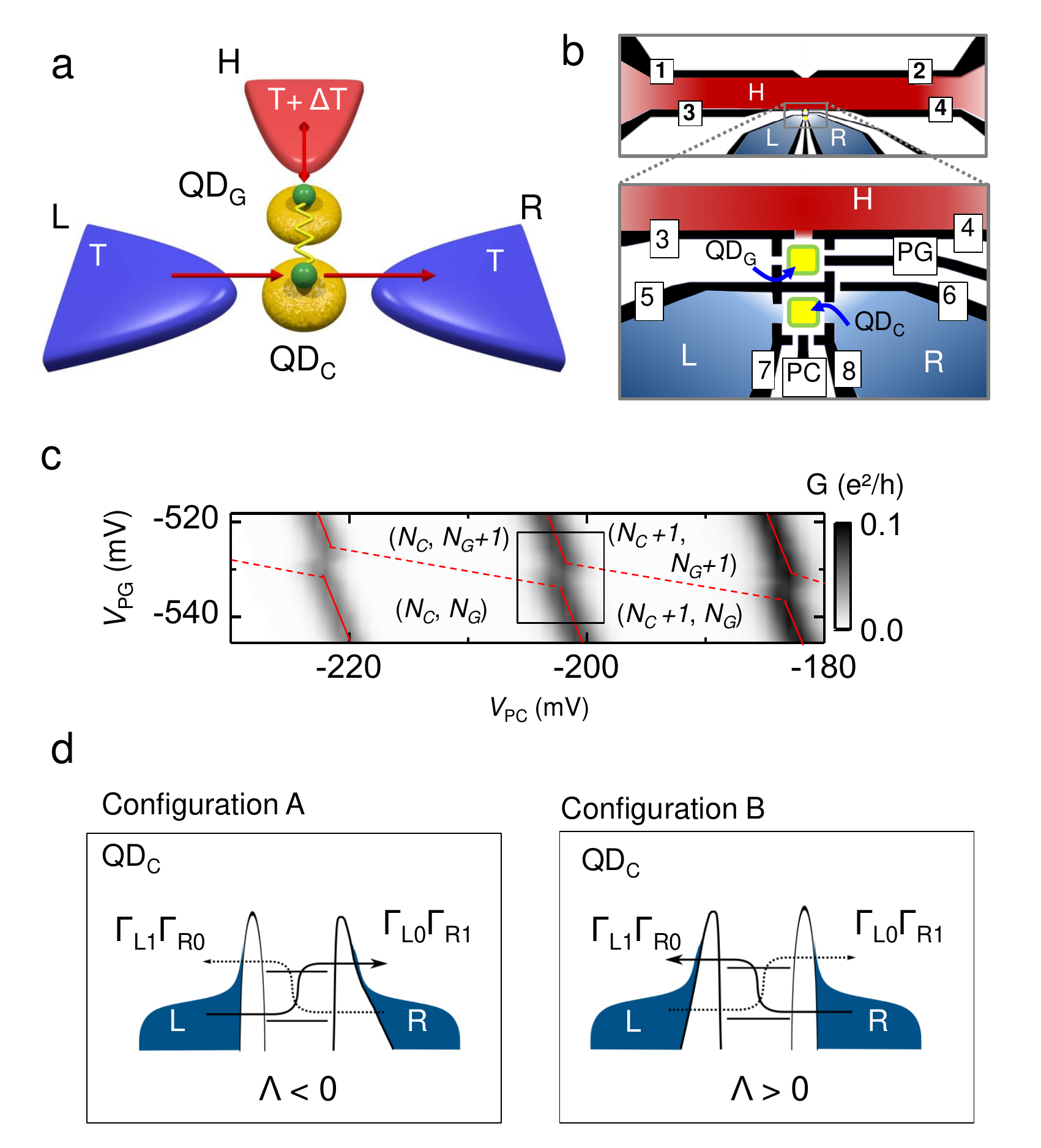}
	\caption{\textbf{a} Schematic of the three-terminal device geometry. One electronic reservoir H (red) is at a high temperature ($T + \Delta T$) and can exchange electrons (green) with quantum dot QD$_{\rm G}$ (yellow disc), as indicated with a red arrow. Two other reservoirs (L and R, blue) at a lower temperature $T$ are tunnel coupled to another quantum dot (QD$_{\rm C}$). The quantum dots interact only capacitively (yellow wave) while particle exchange between them is suppressed. \textbf{b} Schematic of the gate layout (black). The individual gates are denoted with numbers. Gates 1-4 form the heating channel, representing reservoir H, gates 3-8 form the QD system. In addition gates PC and PG are used to tune the energies of the dots individually. \textbf{c} Conductance ($G$) stability diagram of the QD system for configuration $A$ and T$_\text{H} = T_\text{L,R}=\unit[230]{mK}$. Borders of the stability regions are denoted with red lines (solid and dashed), regions of stable charge configuration are labeled ($N_{\rm C}, N_{\rm G}$). \textbf{d} Cartoon sketches of QD$_{\rm C}$ for two configurations $A,B$ of tunnel barrier settings each corresponding to different products of tunneling coefficients $\Gamma_{\rm L0} \Gamma_{\rm R1}$ and $\Gamma_{\rm L1} \Gamma_{\rm R0}$ which leads to different signs for $\Lambda$.
	}
	\label{fig:2}
\end{figure}

Here we present the first experimental realization of a three-terminal thermoelectric energy harvester operating at the nanoscale. We realize a device proposed recently \cite{Sanchez_Optimal_2011} using the layout sketched in Fig.~\ref{fig:2}a: One terminal (H) serves as a heat reservoir at a high temperature $T + \Delta T$. The other two terminals (L,R) are kept at a lower temperature $T$ and constitute the conductor circuit. Particle exchange between these two subsystems of different temperature is suppressed. Energy transfer, however, is provided through a two-quantum dot system (QD$_{\rm C}$, QD$_{\rm G}$) which is tunnel coupled to the heat reservoir on one side through QD$_{\rm G}$ and to both reservoirs of the conductor system on the other side through QD$_{\rm C}$. The two dots interact with each other only capacitively. Hence, a change in occupation number ($N_{\rm C}$, $N_{\rm G}$) by one on either dot changes the energy of the respective other dot by an amount $U$ equal to the electrostatic coupling energy \cite{Molenkamp_Scaling_1995}. Because particle exchange with the electronic reservoirs is enabled for each dot individually, the occupation numbers, and thus energies, of the dots fluctuate according to the thermal fluctuations of carriers in the reservoirs. 
In order to harvest these thermal fluctuations our device uses the working principle depicted in the cycle of energy diagrams in Fig.~\ref{fig:1}. When occupation fluctuations on both QDs are correlated such that the QD system undergoes the sequence $(N_\text{C},N_\text{G})\to(N_\text{C}+1,N_\text{G})\to(N_\text{C}+1,N_\text{G}+1)\to(N_\text{C},N_\text{G}+1)\to(N_\text{C},N_\text{G})$ an energy amount $U$ is extracted from reservoir H and is delivered to the cold subsystem. This defines the direction of heat flow $J$. If the cold conductor system is symmetric with respect to L and R, the energy is dissipated equally to either lead on a time average. However, we can use the tunneling rates $\Gamma_{{\rm L},i}$ and $\Gamma_{{\rm R},i}$ of the barriers at low ($i=0$) and high ($i=1$) energies to break this symmetry. In this case energy flow becomes directed: If the tunneling coefficients are asymmetric with respect to both energy and reservoirs, electrons intrinsically favor entering QD$_{\rm C}$ from one reservoir at a low energy and leaving it to the other reservoir at a higher energy.    
This generates a charge current which, in an unbiased conductor, is proportional to the asymmetry factor $\Lambda=(\Gamma_{\rm L0}\Gamma_{\rm R1}-\Gamma_{\rm L1}\Gamma_{\rm R0})/\Gamma_{0}\Gamma_{1}$, with $\Gamma_{i}=\Gamma_{{\rm L}i}+\Gamma_{{\rm R}i}$, and the transferred heat current, $J$,~\cite{Sanchez_Optimal_2011}
\begin{equation}
I=\frac{e\Lambda}{U}J.
\label{Eq:1}
\end{equation}
Hence, for a given $J$ the direction of the generated current depends only on the sign of $\Lambda$.

For the experimental realization it is therefore crucial to gain direct control over the tunneling rates.
For this purpose top-gate defined quantum dot structures are especially well suited because they allow for a high flexibility in tuning the coupling energy of a QD to its environment.
Our device is realized with this technique using the pattern of gate-electrodes shown in Fig.~\ref{fig:2}b on a GaAs/AlGaAs interface 2DEG.
The two quantum dots QD$_\text{G}$ and QD$_\text{C}$ are positioned in close vicinity to each other in order to ensure a sufficiently large $U$ through good capacitive interdot coupling. 
We use the electrodes PC and PG to control the electro-chemical potential $\mu_\text{C}$ and $\mu_\text{G}$ of QD$_\text{C}$ and QD$_\text{G}$, respectively. This allows us to precisely adjust the electron occupation number $(N_\text{C},N_\text{G})$ of the system. 
The voltages applied to gates 7 and 8 directly influence the corresponding tunneling barriers. In this manner we can control the tunneling asymmetry $\Lambda$ in the conductor system:  
A state of broken left-right symmetry is obtained by increasing $V_8$ so that the potential barrier height between QD$_{\rm C}$ and reservoir R is increased thus reducing the corresponding tunneling coefficients.
Energy dependent tunneling rates typically occur quite naturally in top gate defined structures \cite{MacLean_Energy_2007}. However, a direct control over this parameter is desirable. Therefore, we apply a smaller bias to gate 6 than to gate 5. This strongly affects the shape of the potential barrier between QD$_\text{C}$ and R due to the change in potential landscape in the vicinity of the tunnel junction. In doing so we emphasize the energy dependence of electron tunneling rates between QD$_\text{C}$ and L or R. A simple sketch of the potential barriers associated with this gate voltage configuration (called configuration $A$ from now on) is given in Fig.~\ref{fig:2}d.
From conductance peak amplitude analysis we determine the asymmetry of tunneling coefficients at the Fermi level $\gamma$ for the left and right barrier individually [cf. supplementary material]. This gives $ \gamma_L  =\unit[32.2]{\mu eV} \approx \gamma_R \times 2.6$.

When we measure the conductance $G$ of QD$_\text{C}$ without a temperature difference applied ($T_\text{H}= T_\text{L,R} \approx \unit[230]{mK}$ electron temperature) while varying $V_\text{PC}$ and $V_\text{PG}$ we obtain the stability diagram shown in Fig.~\ref{fig:2}c. Here high $G$ corresponds to an alignment of $\mu_{\rm C}$ with $\mu_{\rm L,R}$, denoted with red, solid lines. Dashed lines indicate configurations for which $\mu_\text{G} = \mu_\text{H}$. The intermediate areas correspond to regions of stable charge configurations.  
At the meeting points of three stability regions (triple points, TP) both QD occupation numbers can fluctuate \cite{VanderWiel_Electron_2002}. The TPs occur in pairs, separated by the interaction energy $U$. From d$I$/d$V$ measurements and the data in Fig.~\ref{fig:2}c we obtain $U \approx \unit[70]{\mu eV}$. For our energy harvester, optimal working conditions exist when for the $(N_{\rm C}, N_{\rm G})$ state, $\mu_{\rm C}$ and $\mu_{\rm G}$ both lie closely below the chemical potential of the adjacent reservoir, while they are situated above when the energy is increased by $U$ (cf. Fig.~\ref{fig:1}). Thus, the region of interest is between two neighboring TPs. 

In order to establish a temperature difference across the QD-system we use a current heating technique \cite{Molenkamp_Quantum_1990} by applying an ac-current of $I_{\rm h} = \unit[150]{nA}$ at $f= \unit[11]{Hz}$ to reservoir H. This increases locally the electron temperature in reservoir H by $\Delta T \approx \unit[100]{mK}$ while the other reservoirs remain at base temperature. Moreover, it ensures that all signals resulting from a change of $T_{\rm H}$ occur at a doubled frequency $2f$, allowing a convenient detection with lock-in technique [see methods].
We now detect the electrons pushed into reservoir R due to a temperature increase $\Delta T$ in reservoir H by using a current amplifier connected to R and a lock-in amplifier detecting at $2f$, while reservoir L is grounded.

\begin{figure}[]
	\centering
	\includegraphics[width=0.99\linewidth]{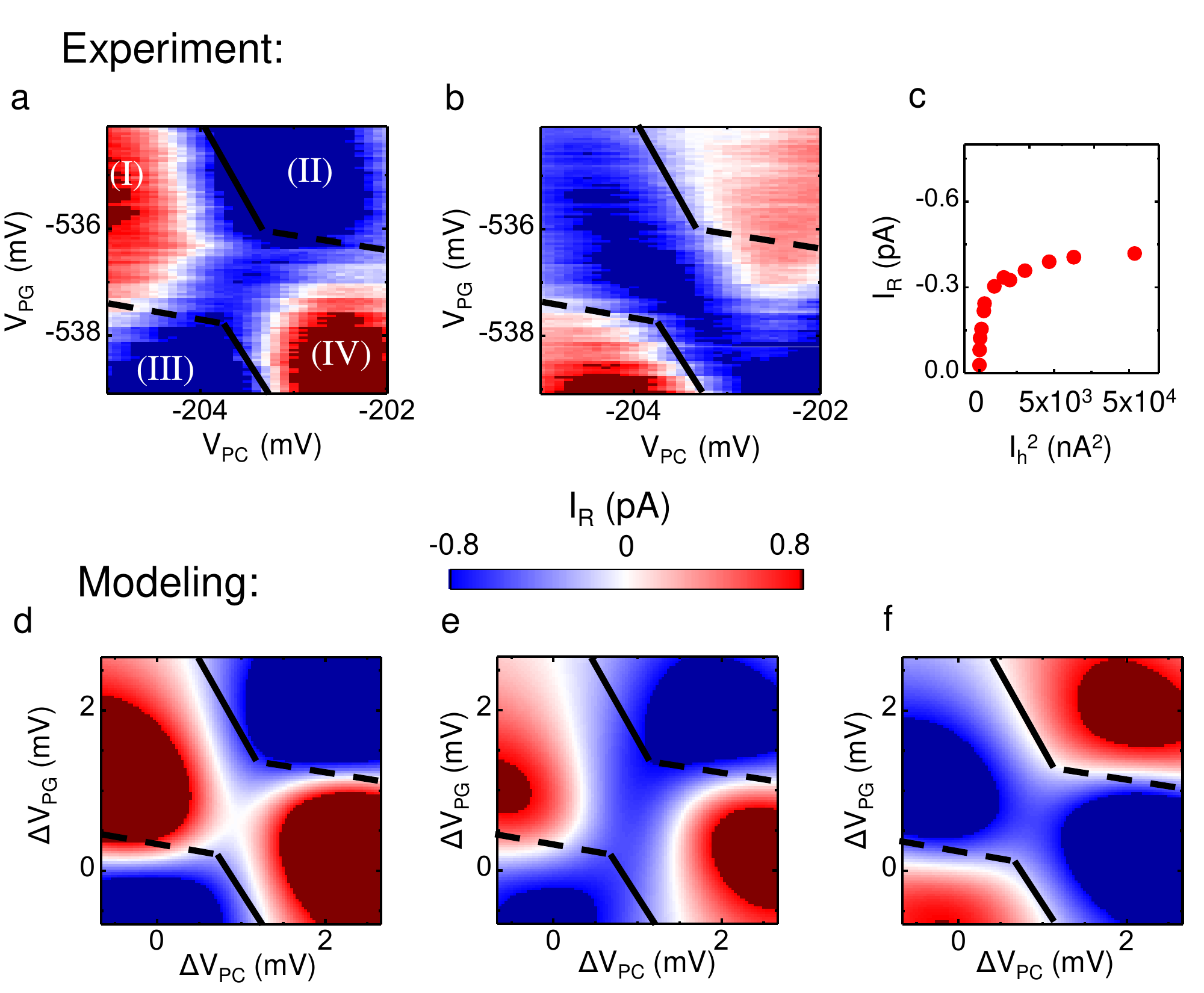}
	\caption{2f-current in reservoir R, $I_{\rm R}$, for configuration $A$ in the vicinity of a TP pair. Black lines denote the stability region borders as obtained from the conductance data. \textbf{a} Experimental data for $0<\Delta \mu_{\rm LR}<\unit[10]{\mu V}$. The signal around the TP pair is a result of thermal gating (regions I - IV). It becomes reversed if $\Delta \mu_{\rm LR}$ is inverted, $-\unit[10]{\mu V}<\Delta \mu_{\rm LR}<0$, as shown in \textbf{b}. The signal between the TPs is due to the proposed mechanism of energy harvesting. It stays negative, irrespective of the sign of the voltage bias $\Delta \mu_{\rm LR}$. \textbf{c} $I_{\rm R}$ as a function of squared heating current $I_{\rm h}^2$ between two TPs for slightly different $\Lambda$. 
		\textbf{d} Model calculation for energy dependent tunneling barriers of QD$_{\rm C}$, symmetric with respect to L and R. The signal between the TP is zero, only the effect of thermal gating is present. \textbf{e} Calculation using asymmetric and energy dependent tunnel barriers as obtained for configuration A with $0<\Delta \mu_{\rm LR}<\unit[10]{\mu V}$. \textbf{f} Model calculations for $-\unit[10]{\mu V}<\Delta \mu_{\rm LR}<0$.}
	\label{fig:3}
\end{figure}

Figure~\ref{fig:3}a shows the detected signal for the TP pair framed in Fig.~\ref{fig:2}c. Black lines denote the borders of the stability regions as obtained from the conductance data. Surrounding the stability vertex we observe a finite current signal of both positive and negative sign (section I-IV in Fig.~\ref{fig:3}a). This signal is due to thermal gating \cite{Thierschmann_Thermal_2015} of a small current resulting from a finite potential difference $\Delta\mu_{\rm LR}$  between reservoirs L and R which is created by the current amplifier [$10~\mu {\rm V} > \Delta \mu_{\rm LR} > 0 $].
This effect is expected to give contributions only outside the TP region. However, between the TPs is where we expect the energy harvester to have its largest output current. In this region a finite negative current $I_{\rm R} \approx -\unit[0.6]{pA}$ is clearly observed. We verify that this current is not due to $\Delta \mu_{\rm LR}$ by reversing the potential difference [Fig.~\ref{fig:3}b]. As expected for thermal gating \cite{Thierschmann_Thermal_2015}, the current outside the stability vertex changes sign. However, at the center, the direction of $I_{\rm R}$ is independent of a small bias voltage. 

Theory predicts $I_{\rm R}$ to first sharply increase with $\Delta T$ followed by a flattening of the curve for large $\Delta T$ \cite{Sanchez_Optimal_2011}. In Fig.~\ref{fig:3}c, we show $I_{\rm R}$ as a function of the squared heating current ($\Delta T \propto I_{\rm h}^2$, cf. methods) for a slightly different $\Lambda$. The observed behavior is in qualitative agreement with theory. 

Our observations are confirmed by model calculations based on the rate equation approach presented in Ref.~\cite{Sanchez_Optimal_2011}, using parameters extracted from the experiments [cf. methods and supplementary material]. For fully symmetric tunneling coefficients ($\Lambda = 0$) the calculations reproduce the current signal in the periphery of the TP pair while zero current is obtained at the center between the TPs (Figs.~\ref{fig:3}d). When introducing a small asymmetry, $\Lambda \neq 0$, heat conversion becomes enabled. A finite current peak appears between the TP, thus yielding excellent agreement with the experiments for both $\Delta \mu_{\rm LR}<0$ (Fig.~\ref{fig:3}e) and $\Delta \mu_{\rm LR}>0$ (Fig.~\ref{fig:3}f). This is strong evidence that $I_{\rm R}$ observed in this region is indeed a result of the conversion of thermal energy into a directed charge current.  

\begin{figure}[]
	\begin{center}
		\includegraphics[width=0.7\linewidth]{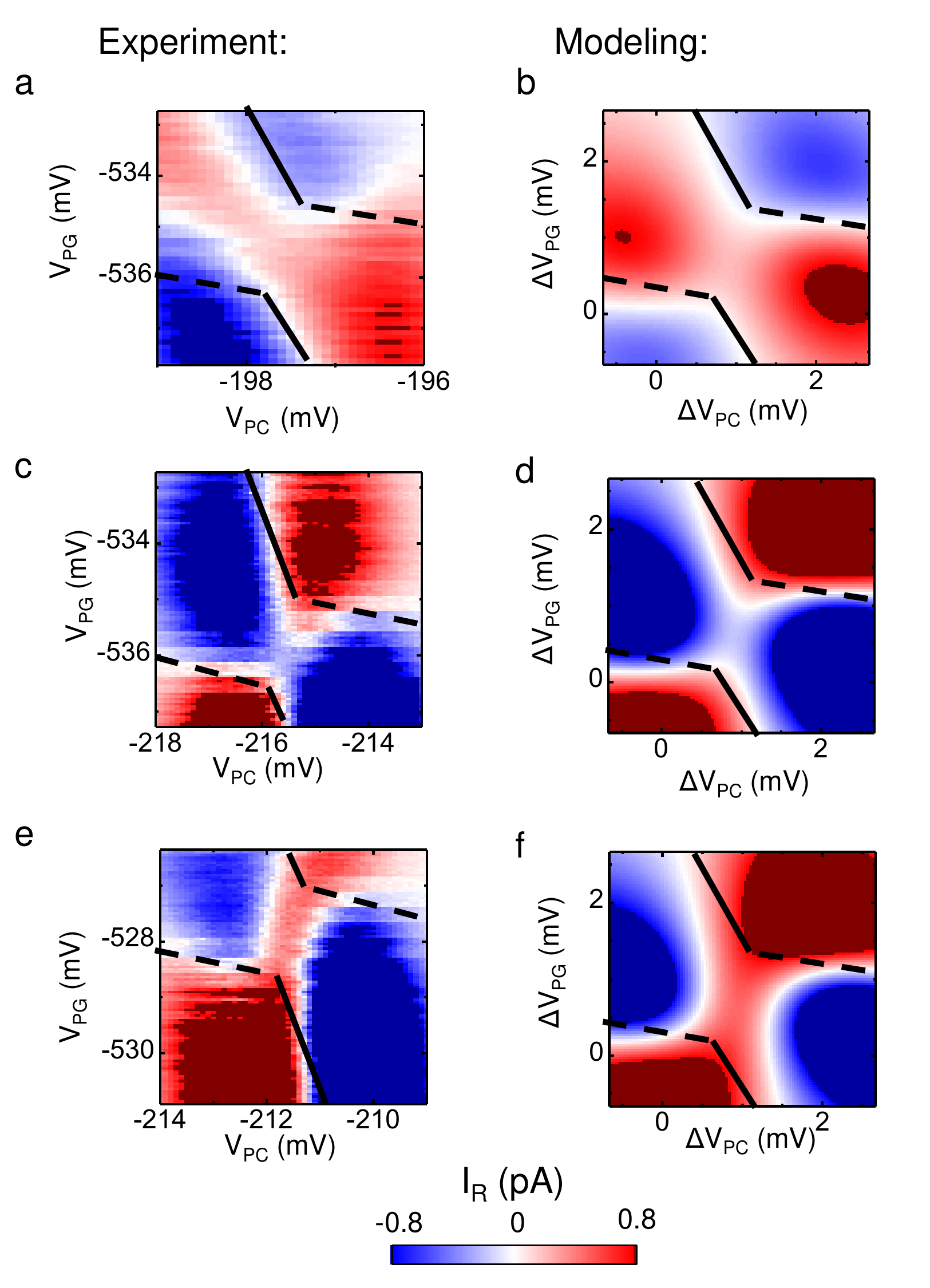}
	\end{center}
	\caption{\textbf{a} Measured $I_{\rm R}$ for configuration $B$. \textbf{b} Model calculation for configuration $B$ \textbf{c} $I_{\rm R}$ for tunnel barriers tuned to symmetric tunneling coefficients at the Fermi level of the reservoirs. A finite $I_{\rm R}$ is still visible at the center due to a difference in energy dependence of the confining potential barriers. Corresponding model calculations are shown in \textbf{d}. \textbf{e} Increasing the gate voltage applied to gate 6 changes the energy dependence of the tunnel barrier connecting QD$_{\rm C}$ to reservoir R. This also inverts the tunneling asymmetry factor $\Lambda$ and thus the generated charge current changes sign. \textbf{f} Model calculation for changed energy dependence of tunneling coefficients. 
	}
	\label{fig:4}
\end{figure}

As mentioned above, a unique property of our energy harvesting device, and a direct result of its three-terminal geometry, is that the direction of the generated charge current $I_{\rm R}$ is determined by the asymmetry factor $\Lambda$ of the tunneling coefficients. In our experiment, this parameter can be controlled by external gate voltages.
When we increase the gate voltage V$_7$ while reducing V$_8$, i.e. pinching off reservoir L more strongly from QD$_{\rm C}$ than reservoir R [cf. supplementary], we obtain potential barriers as shown in the sketch in Fig.~\ref{fig:2}d labeled configuration $B$. This setting corresponds to an inverted left-right asymmetry compared to configuration $A$. Because the energy dependence of the tunneling barriers is still present, we can expect $\Lambda$ to be inverted, as well. 
The resulting $I_\text{R}$ is shown in Fig.~\ref{fig:4}a. We now observe a positive $I_\text{R} \approx \unit[0.2]{pA}$ between the TP. 
This directly demonstrates how the direction of the thermally generated current can be manipulated by inverting the asymmetry of the tunneling rates between QD$_{\rm C}$ and L, R. 
This is also consistent with model calculations which are based on the experimental parameters for configuration $B$ [cf. Fig.~\ref{fig:4}b].

A similar effect can be achieved if we manipulate the energy dependence of the barriers. Figure~\ref{fig:4}c shows data for a barrier configuration which exhibits symmetric tunneling coefficients at the Fermi level. [Note, that for these measurements $-\unit[10]{\mu V} < \Delta \mu_{\rm LR} < 0$.] However, since the barriers exhibit different shapes, $\Lambda \neq 0$. As shown in Fig.~\ref{fig:4}e the asymmetry can be tuned by carefully increasing the voltage applied to gate 6. This mainly affects the shape and thus the energy dependence of the tunneling barrier connecting QD$_{\rm C}$ and R. As a result the thermally generated current changes the direction. Again, the experiments are in excellent agreement with theory [cf. Figs.~\ref{fig:4}d, f].  

As discussed in detail in Ref.~\cite{Sanchez_Optimal_2011}, energy harvesters based on our mechanism can, in principle, work as an optimal heat engine reaching Carnot efficiency $\eta_C$. The device performance is directly related to the tunneling asymmetry $\Lambda$, which is inherently low in gate defined quantum dots. We obtain excellent agreement with the experiments when we use $\Lambda = 0.04$ for configuration A and $\Lambda = 0.01$ for configuration B in the model calculations. We point out that a further increase of $\eta$ can be achieved by an optimized sample layout which improves the tunneling asymmetry $\Lambda$ (e.g. by a more elaborate injection scheme \cite{Sanchez_Optimal_2011}) and also the achievable temperature difference $\Delta T$ and the parameter $U$. 

\subsection{Methods}

\textit{Heating Current Technique:}
The heating channel, reservoir H, has a width of $\unit[2]{\mu m}$ over a length of $\unit[20]{\mu m}$, shaped by gates 1-4. The constriction at its center formed by gates 1 and 2 can be used as a voltage probe for the channel. By adjusting $V_1$ and $V_2$, its conductance is set to $G=\unit[10]{e^2/h}$ thus ensuring that no thermovoltage arises when the temperature in reservoir H is increased. Left and right of the channel, reservoir H opens into large contact areas. 
The temperature $T_{\rm H}$ in the heat bath H is controlled by applying an ac-current $I_{\rm h} = \unit[150]{nA}$ at a frequency of $f = \unit[11]{Hz}$ to the channel \cite{Molenkamp_Quantum_1990}. This introduces the Joule-heating-power $P \propto I_{\rm h}^2$ into the electron gas. 
Due to the reduced electron-lattice interaction in GaAs/AlGaAs 2DEGs at low temperature, electron-electron scattering dominates electron-phonon scattering in the channel. Thus, the electron temperature increases here by $\Delta T$. In the wide contact areas left and right of the channel, however, spatial dimensions are sufficiently large to ensure cooling of the 2DEG by electron-lattice scattering. Hence, we locally increase the temperature in the channel by $\Delta T \approx \unit[100]{mK}$ (estimated from QPC thermometry, \cite{Molenkamp_Quantum_1990}). The quadratic relation between $I_{\rm h}$ and $P$ causes $T_{\rm H}$ to oscillate with twice the frequency of I$_{\rm h}$. This provides thermal effects in the device with the signature of an oscillation frequency $2f = \unit[22]{Hz}$.

\textit{Theory:}
For the calculations we use a rate equation approach as presented in the model section of the supplementary material. 
The following parameters are used. $\mu_{\rm L,R}=\pm\unit[5]{\mu V}$, $U=\unit[70]{\mu eV}$, $T=\unit[19.82]{\mu eV}$, $\Delta T=\unit[8.6]{\mu eV}$. The tunneling rates for QD$_{\rm G}$ have been chosen to be: $\Gamma_{\rm H0}=\Gamma_{\rm H1}=\unit[10.0]{\mu eV}$.
In order to obtain the results shown in Fig.~\ref{fig:3}d we use symmetric tunneling coefficients $\Gamma_{\rm L0} = \Gamma_{\rm R0} = \unit[32.2]{\mu eV}$, $\Gamma_{\rm L1} = \Gamma_{\rm R1} = \unit[25.8]{\mu eV}$. 
The tunneling rates used for configuration $A$ in Fig.~\ref{fig:3}e and f are $\Gamma_{\rm L0}=\unit[23.2]{\mu eV}$, $\Gamma_{\rm L1}=\unit[25.8]{\mu eV}$, $\Gamma_{\rm R0}=\unit[9.0]{\mu eV}$, $\Gamma_{\rm R1}=\unit[12.5]{\mu eV}$. For Configuration $B$ we have used (Fig.~\ref{fig:4}): $\Gamma_{\rm L0}=\unit[4.6]{\mu eV}$, $\Gamma_{\rm L1}=\unit[5.5]{\mu eV}$, $\Gamma_{\rm R0}=\unit[23.0]{\mu eV}$, $\Gamma_{\rm R1}=\unit[24.2]{\mu eV}$.
The behaviour observed in Fig.~\ref{fig:4}c is reproduced by the model, when we allow the tunneling rates to have a small energy dependence around the Fermi energy, which is different in each barrier: i.e. $\gamma_l\approx(\Gamma_{l0}+\Gamma_{l1})/2$, and $\Delta\Gamma_\text{L}\ne\Delta\Gamma_\text{R}$ with $\Delta\Gamma_l=\Gamma_{l0}-\Gamma_{l1}$ and $\gamma_\text{L}=\gamma_\text{R}$. (Parameters used: $\gamma_l=(\Gamma_{l0}+\Gamma_{l1})/2$: $\Gamma_{\rm L0}=\unit[21.6]{\mu eV}$, $\Gamma_{\rm L1}=\unit[24.9]{\mu eV}$, $\Gamma_{\rm R0}=\unit[20.9]{\mu eV}$, $\Gamma_{\rm R1}=\unit[25.5]{\mu eV}$)
This is enough to generate a response (cf. Fig~\ref{fig:4}d) which is then tunable by increasing the energy dependence in one of the barriers, in this case $\Delta\Gamma_{\rm R}$, while keeping the average $\gamma_\text{R}$ fixed (cf. Fig~\ref{fig:4}f) ($\Gamma_{\rm L0}=\unit[21.0]{\mu eV}$, $\Gamma_{\rm L1}=\unit[25.4]{\mu eV}$, $\Gamma_{\rm R0}=\unit[20.3]{\mu eV}$, $\Gamma_{\rm R1}=\unit[21.2]{\mu eV}$).

\subsection{Acknowledgements}
We gratefully acknowledge Markus B\"uttiker for drawing our attention to the subject. Furthermore, the authors thank Cornelius Thienel for discussions and Luis Maier for help with device fabrication. 
This work has been supported by the Deutsche Forschungsgemeinschaft via SPP1386, the Swiss National Science Foundation, the Spanish MICINN Juan de la Cierva program and MAT2011-24331, COST Action MP1209.

\subsection{Author Contributions}

H.T., H.B. and L.W.M. designed the experiment, C.H. and W.H. provided the wafer material, F.A. fabricated the sample, H.T. and F.A. conducted the measurements, B.S. and R.S. performed the model calculations. All authors discussed the results. H.T., B.S., R.S., H.B. and L.W.M. wrote the manuscript.

\subsection{Competing financial interests}
The authors declare no competing financial interests.

\newpage

\section{Supplementary}

\subsection{Sample Information and Fabrication Technique}

The device is fabricated from a GaAs/AlGaAs wafer containing a 2-dimensional electron system (2DES) at 92 nm below the surface. The nominal carrier density $n$ and mobility $\mu$ at 4K are $n =-\unit[2.14 \times 10^{11}]{cm^{-2}}$ and $\mu = \unit[0.71 \times 10^{6}]{cm^{2}(Vs)^{-1}}$. The electronic reservoirs are equipped with annealed Au/Ge pads which provide good electrical contact to the 2DES. Using standard optical and e-beam lithography Ti/Au electrodes are patterned onto the sample surface which can then be used as gates to deplete the 2DES underneath by applying negative voltages with respect to the electron system.

\begin{figure}
	\centering
	\includegraphics[width=0.85\linewidth]{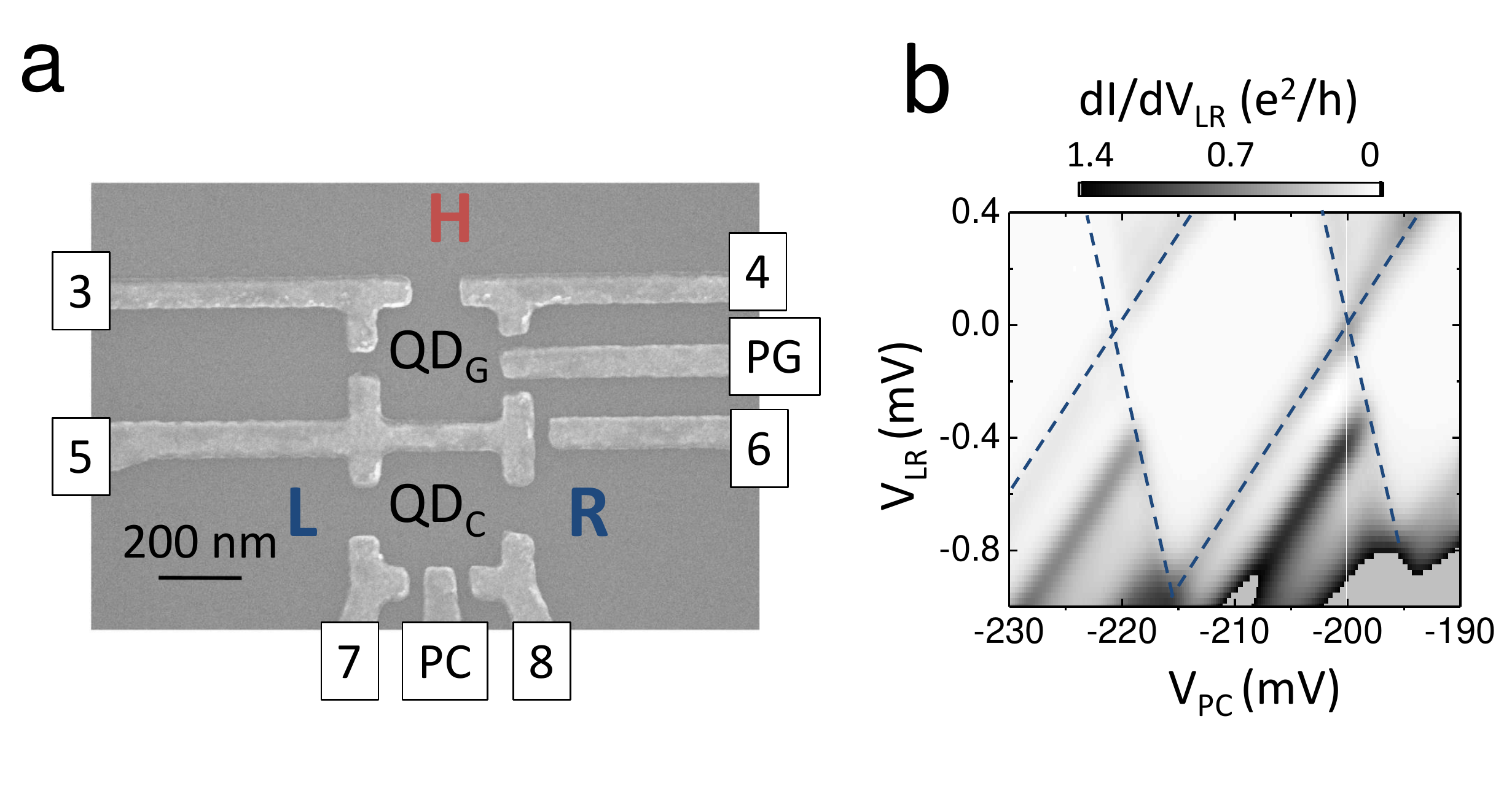}
	\caption{\textbf{a} SEM-micrograph of the QD system \textbf{b} d$I$/d$V$ data obtained for QD$_{\rm C}$. Dashed lines indicate the Coulomb diamonds.}
	\label{fig:S1}
\end{figure}

An SEM-micrograph of the QD-system is given in Fig.\ref{fig:S1} a. The lithographical size of each QD is $\unit[(300 \times 300)]{nm^2}$. However, due to electrostatic depletion by the gates the effective QD size is expected to further decrease to approximately $\unit[(200 \times 200)]{nm^2}$. From these numbers and the electron density $n$ an average QD occupation number of the order of 100 can be estimated. d$I$/d$V$ measurements for QD$_{\rm C}$ [cf. Fig.~\ref{fig:S1} b] yield a single dot charging energy of $E_{\rm C} = \unit[0.95]{meV}$.

\subsection{Tunnel Barrier Asymmetry}
\begin{figure}
	\centering
	\includegraphics[width=0.85\linewidth]{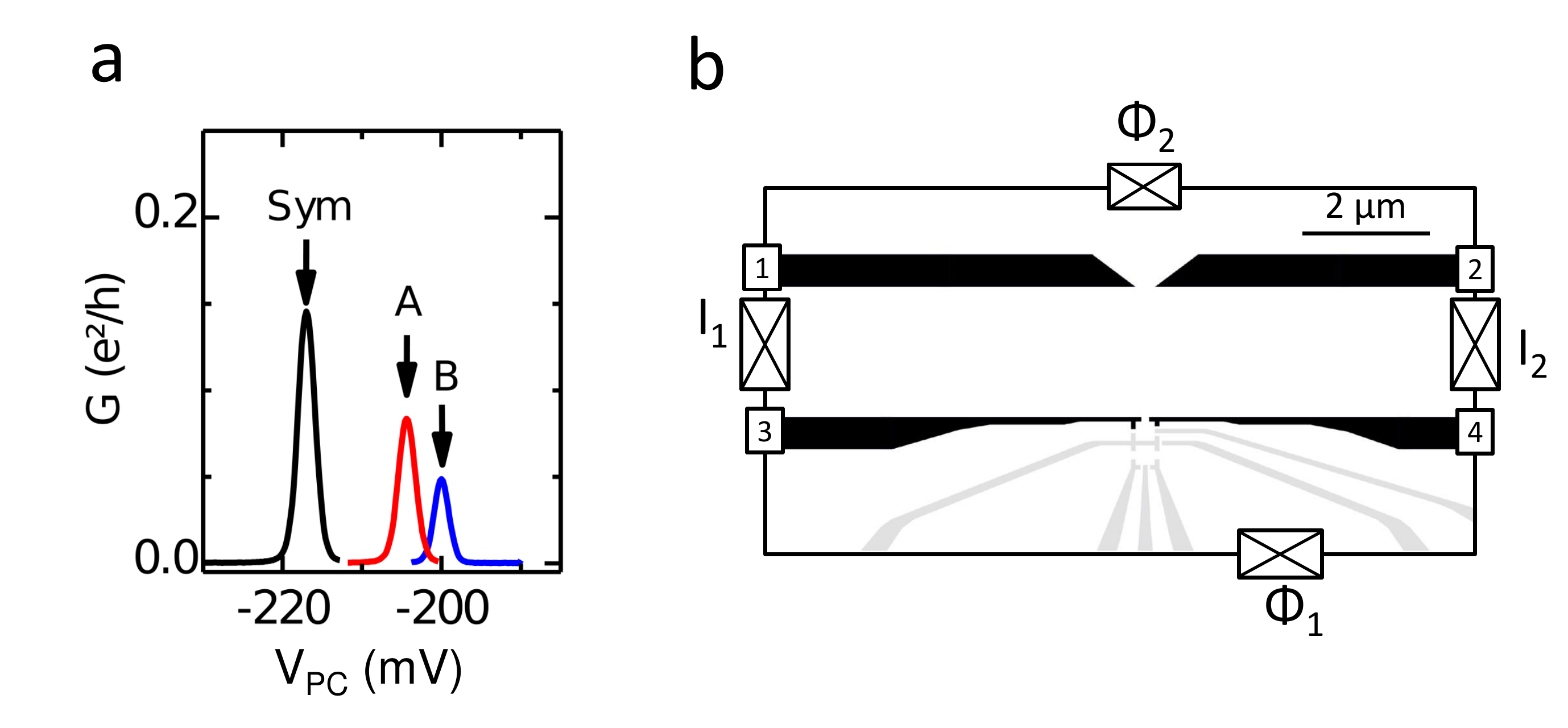}
	\caption{\textbf{a}  Coulomb conductance resonance of QD$_{\rm C}$ for gate configurations with $\gamma_{\rm L} = \gamma_{\rm R}$ ($Sym$, black), $\gamma_{\rm L} = 2.6 \gamma_{\rm R} $ ($A$, red) and $\gamma_{\rm R} = 5\gamma_{\rm L}$ ($B$, blue). \textbf{b} Schematic of the gate geometry. For QPC thermometry only those gates which define the heating channel are energized (black). All other gates are grounded (gray). White indicates conducting regions. The heating current is applied between contacts $I_{\rm 1}$ and $I_{\rm 2}$. The thermovoltage $V_{\rm T}$ is detected between $\Phi_{\rm 1}$ and $\Phi_{\rm 2}$.} 
	\label{fig:S2}
\end{figure}

The conductance amplitude $G_0$ of a Coulomb resonance is given by 

\begin{equation}
G_0 = \frac{1}{4 k_B T} \frac{\gamma_{\rm L} \gamma_{\rm R}}{\gamma_{\rm L}+\gamma_{\rm R}} \frac{e^2}{h}
\end{equation}

where $\gamma_L$ and $\gamma_R$ denote the tunneling coefficient at the Fermi level between QD$_{\rm C}$ and reservoirs L and R, respectively. Obviously, Eq.~1 becomes largest for symmetric barriers, i.e. $\gamma_L = \gamma_R$. This relation can be used to obtain quantitative information about the coefficients for tunneling processes between QD$_{\rm C}$ and reservoirs L and R in our experiment. We therefore proceed as follows:
First, $V_7$ and $V_8$ [cf. Fig.~\ref{fig:S1}a] are tuned so that the amplitude of the conductance peak is maximized ($V_8 = \unit[614]{mV}$, $V_7 = \unit[520]{mV}$). This gate voltage configuration then corresponds to symmetric coupling energies at the Fermi level $E_{\rm F}$. The conductance peak obtained for this setting is labeled \textit{Sym} (black) in Fig.~\ref{fig:S2}a. For the peak amplitude we obtain $G_0 = \unit[0.145]{e^2/h}$. Using Eq.~1 and $T = \unit[230]{mK}$ then yields $\gamma_{\rm L}  = \gamma_{\rm R} = \unit[23.2]{\mu eV}$. 
Next, we carefully increase $V_8$ by $\unit[15]{mV}$.

This causes the conductance peak amplitude to decrease. This new gate voltage setting corresponds to configuration $A$. The conductance resonance for this configuration is shown in Fig.~\ref{fig:S2}a (red). We determine the peak value to be $G_0 = \unit[0.084]{e^2/h}$. As a first approximation we assume that a variation of $V_8$ only affects $\gamma_{\rm R}$ while $\gamma_{\rm L}$ stays constant ($\gamma_{\rm L}= \unit[23.2]{\mu eV}$). In this case, we obtain $\gamma_{\rm R}= \unit[9]{\mu eV}$. This gives for the source-drain asymmetry $\gamma_{\rm L} \approx 2.6 \gamma_{\rm R}$. 
In a similar manner we can proceed with the left barrier. When $V_7$ is increased instead of $V_8$ by \unit[16]{mV}, we obtain the opposite barrier asymmetry, configuration $B$. For this setting we obtain the peak denoted $B$ in Fig.~\ref{fig:S1}b (blue) with $G_0 = \unit[0.048]{e^2/h}$. Correspondingly, we find $\gamma_{\rm L} = \unit[4.6]{\mu eV}$ while $\gamma_{\rm R}= \unit[23.2]{\mu eV}$. Thus, the tunneling coefficients at the Fermi level relate to each other as $\gamma_{\rm R} \approx 5 \gamma_{\rm L}$.

\subsection{Quantum Point Contact Thermometry}

In order to estimate the temperature difference arising between the heating channel (reservoir H) and the other reservoirs of the device, we use Quantum Point Contact (QPC) thermometry \cite{Streda_Quantized_1989, Molenkamp_Quantum_1990, Houten_Thermo_1992, Appleyard_Thermometer_1998}.

\begin{figure}
	\centering
	\includegraphics[width=0.85\linewidth]{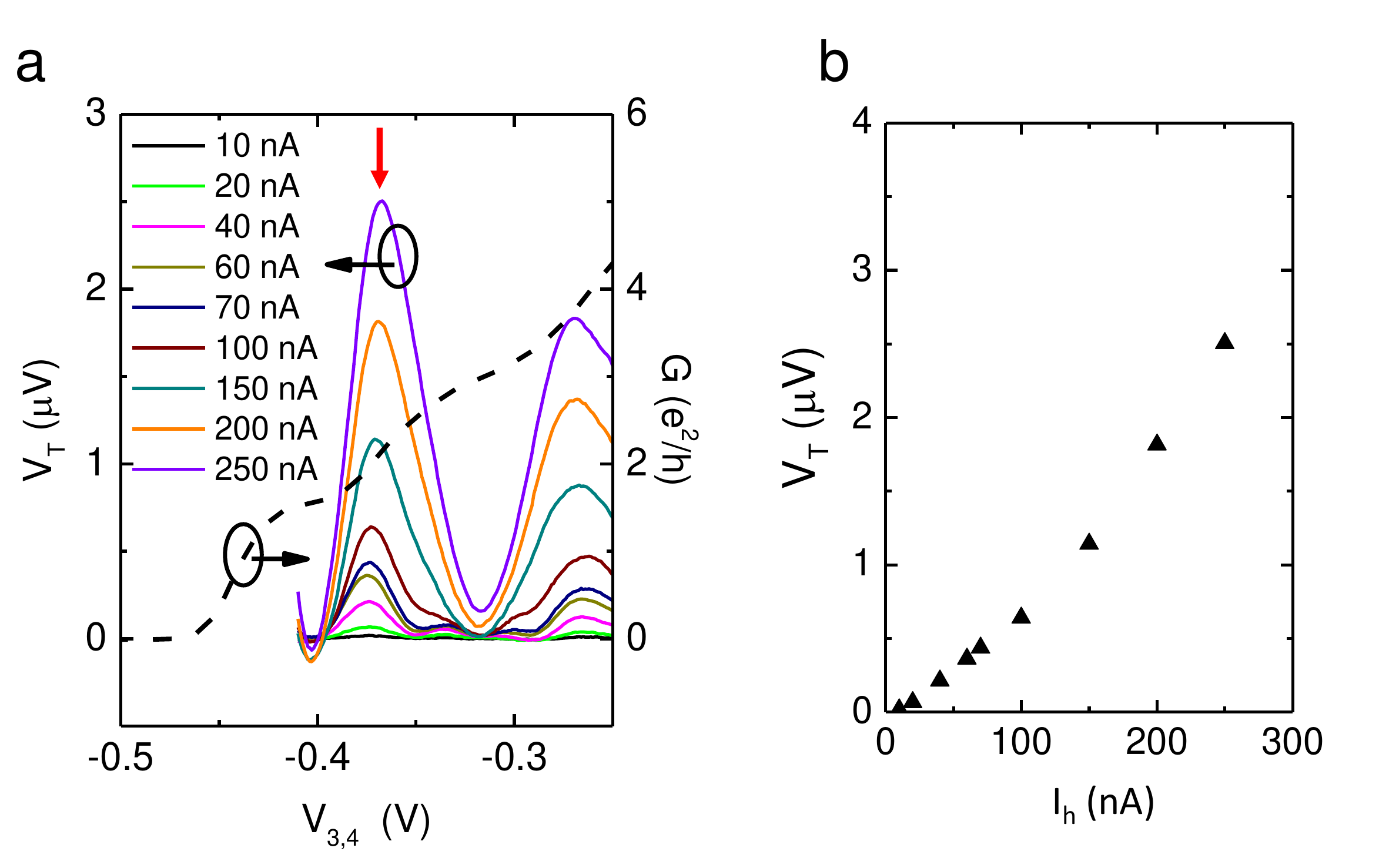}
	\caption{\textbf{a} Thermovoltage $V_{\rm T}$ of QPC 3,4 for $I_{\rm h}=\unit[10]{nA}...\unit[250]{nA}$ with QPC 1,2 set to the $G=\unit[10]{e^2/h}$ plateau. The dashed line shows the measured conductance (right axis). \textbf{b} $V_{\rm T}$ vs. $I_{\rm h}$ extracted from the maximum indicated with an arrow in a.}
	\label{fig:S3}
\end{figure}

For the channel temperature measurements all gates are grounded execpt for those which define the heating channel (gates 1,2,3 and 4, cf. Fig. \ref{fig:S2}b). QPC 1,2 is set to the conductance plateau $G=\unit[10]{e^2/h}$, thus ensuring that no thermovoltage arises from this QPC. Next, a heating current $I_{\rm h}$ is applied to the heating channel via contacts $I_{\rm 1}$ and $I_{\rm 2}$ which oscillates with the frequency $f = \unit[11]{Hz}$. Note that $\Phi_{\rm 2}$ is directly connected to ground potential thus keeping the potential at the channel center fixed. The excitation voltage for the heating current is applied in such a manner that the chemical potential at both channel contacts then oscillates symmetrically with respect ot this point. This way it is ensured that no unwanted $2f$ contributions arise from direct capacitive coupling of the heating channel to the QD system.  We then detect the voltage arising between the voltage probes $\Phi_{\rm 1}$ and $\Phi_{\rm 2}$ at the second harmonic $2f=\unit[22]{Hz}$ with a lock-in amplifier while varying the potential of gates 3 and 4. The detected signal is then given by $V_{\rm T} = (S_{\rm QPC 12} - S_{\rm QPC 34})\Delta T$. Because $S_{\rm QPC12}=0$, $V_{\rm T}$ can be assigned entirely to the QPC 3,4. Since for a QPC with a saddle shaped potential $S_{\rm max}\approx 20 \mu V K^{-1}$, $V_{\rm T}$ can be used to calculated the corresponding $\Delta T$.
The detected $V_{\rm T}$ is shown for several heating currents in Fig.~\ref{fig:S3}a. The corresponding conductance data are shown as a dashed line. Fig~\ref{fig:S3}b gives the thermovoltage peak values at the transition from 4 to 2 conducting modes [marked with a red arrow in Fig.~\ref{fig:S3}a] as a function of $I_{\rm h}$. The non-linear but parabolic behavior is evidence that the signal indeed originates from electron heating \cite{Molenkamp_Quantum_1990}. For $I_{\rm h}=\unit[150]{nA}$ we obtain $V_T=\unit[1.15]{\mu V}$ which corresponds to $\Delta T \approx \unit[60]{mK}$. We point out, however, that this number provides only a lower boundary. The steps in QPC conductance are strongly smeared out. This is especially pronounced at the transition from 4 to 2 conducting modes, indicating that the assumption of an ideally saddle shaped potential, as required for $S_{\rm max} \approx \unit[20]{\mu V K^{-1}}$, is only a very rough estimate. $S_\text{max}$ is actually somewhat smaller. In practice, $S$ is therefore usually calibrated from current- and temperature dependent resistance measurements on the channel (Shubnikov-de Haas oscillations or universal conductance fluctuations). Based on this experience, for the present case we estimate $S_{\rm max} \approx \unit[(10-15)]{\mu VK^{-1}}$, thus suggesting that $\Delta T$ actually is of the order of 100 mK.

\subsection{Model Calculations}

In the sequential tunneling regime, transport can be described within a rate equation approach. The vector $\mathbf P=(P_{00},P_{10},P_{01},P_{11})$ contains the probabilities to find the double dot empty, occupied with one electron on QD$_{\rm C}$ and QD$_{\rm G}$, respectively, and occupied with one electron on each dot. 
In the stationary state, the occupation probabilities obey the master equation $\mathbf{W P}=0$ where the transition rates $\mathbf W$ are given by Fermi's golden rule.
In particular, we have $W^\pm_{\alpha n}=\Gamma_{\alpha n} f^\pm_r(\mu_{\alpha n})$ for the rate of an electron tunneling in or out of QD$_{\alpha}$ when the other quantum dot is occupied with $n$ electrons. Here, $f^\pm_r(x)=\{\exp[(x-\mu_r)/k_{\rm B} T_r]+1\}^{-1}$ denotes the Fermi function of reservoir $r$. The addition energies are given by $\mu_{\alpha 0}=\varepsilon_\alpha$ and $\mu_{\alpha 1}=\varepsilon_\alpha+U$. The current through the system is given by $\Tr{\mathbf{W}^I \mathbf{P}}$ where the current rates $\mathbf{W}^I$ take into account the number of electrons transferred during a tunneling event.
The dependence of the level positions on gate voltages is modeled as $\varepsilon_\text{C,G}=\varepsilon_\text{C,G}^0+aeV_{\text{PC,PG}}+beV_{\text{PG,PC}}$ with $a=0.03$ and $b=0.2a$ depending on the geometric capacitances of the system.

For modeling the experimental data we fix the asymmetry at the Fermi level according to the rates determined in the experiment and add a small energy dependence of the barrier transparency. The effect of the load is taken into account as a reference current. The measured signal is given by the difference of the current with and without a temperature gradient $\Delta T$ applied to terminal H: $\Delta I_\text{R}=I_\text{R}(\Delta \mu_{\rm L,R},\Delta T)-I_\text{R}(\Delta \mu_{\rm L,R},0)$.

\end{document}